\documentclass[10pt,conference]{IEEEtran}

\usepackage[T1]{fontenc}
\usepackage[utf8]{inputenc}
\usepackage{graphicx}
\usepackage{amsmath}
\usepackage{amssymb}
\usepackage{amsfonts}
\usepackage{bm}
\usepackage{cite}
\usepackage{booktabs}
\usepackage{array}
\usepackage{multirow}
\usepackage{url}
\usepackage{xcolor}
\usepackage{algorithm}
\usepackage{balance}
\usepackage{microtype}
\usepackage{algpseudocode}

\newcommand{\magma}{\textsc{MAGMA}}

\title{\magma: Mixture-Model Adaptive Gaussian
Model Acceleration\\
      }

\author{
\IEEEauthorblockN{Peter Forcha}
\IEEEauthorblockA{peter.forcha@ufl.edu}
\and
\IEEEauthorblockN{H. Kajekusumadhar}
\IEEEauthorblockA{h.kajekusumadhar@ufl.edu}
\and
\IEEEauthorblockN{Mbua Peter}
\IEEEauthorblockA{mbuapete@gmail.com}
\and
\IEEEauthorblockN{Muhammed Kawser}
\IEEEauthorblockA{muhammedkawser@gmail.com}
\and
\IEEEauthorblockN{Audrey Cyriell Mo}
\IEEEauthorblockA{audreycyriell.mo@ufl.edu}
}

\begin{document}
\maketitle

\begin{abstract}
Conventional FPGA-based Gaussian Mixture Model (GMM) accelerators use offline-trained, fixed parameters, limiting their ability to adapt to evolving scene statistics in long-lived edge systems. We present MAGMA, a fully synthesizable fixed-point FPGA architecture that performs concurrent GMM inference and online Expectation-Maximization (EM) parameter adaptation from a streaming RGB pixel input. MAGMA combines a pipelined inference datapath with a background update engine using hardware-friendly transcendental approximations---a range-reduced Chebyshev exponential, a CLZ-based logarithm, and a shift-and-subtract divider---alongside guards against variance collapse and cluster death that stabilize online fixed-point EM. Implemented on an AMD Spartan-7 XC7S50 with $K=4$ clusters, MAGMA runs at 74.49~MHz using 7,779 LUTs, 91 DSPs, and no block RAM, consuming 274~mW. It achieves an $11.8\times$ inference speedup and an $81\times$ M-step speedup over software, while spatial subsampling reduces per-update pixel volume by $40\times$ with minimal impact on EM convergence. Under a synthetic non-stationary scene, MAGMA's online adaptation improves mean pixel accuracy over a static baseline (81.5\% vs.\ 79.7\%), demonstrating that full online GMM learning is achievable on a commodity edge FPGA.
\end{abstract}

\section{Introduction}
\label{sec:intro}

\begin{figure}[t]
\centering
\includegraphics[width=\columnwidth]{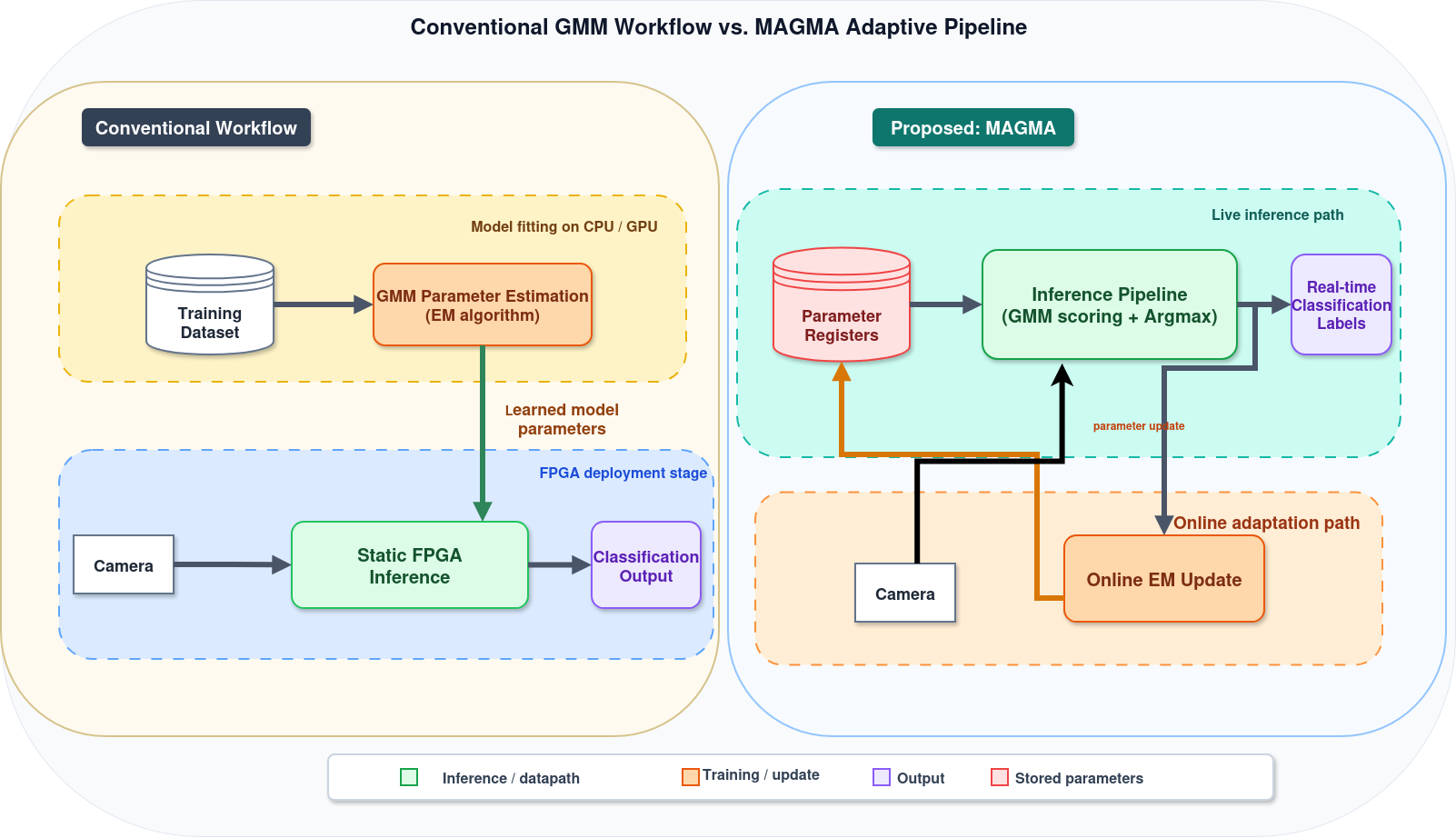}
\caption{Conventional offline-trained GMM accelerators
versus the proposed MAGMA architecture with concurrent
online parameter adaptation.}
\label{fig:magma_comparison}
\vspace{-4mm}
\end{figure}

Many real-world vision systems operate in dynamic environments
where illumination changes, shadows move, and objects enter or
leave the scene. These variations alter the statistical
distribution of pixel intensities over time, requiring models
that can adapt continuously to evolving visual data.
However, most deployed vision pipelines rely on models trained
offline and fixed at deployment time (Fig.~\ref{fig:magma_comparison}). As scene statistics drift,
segmentation accuracy degrades and the conventional remedy is offline retraining and redeployment,
an impractical workflow for always-on edge systems\cite{bayram2022concept}. 
Gaussian Mixture Models (GMMs) are widely used for modelling
visual distributions in applications including image
segmentation, background subtraction, and anomaly detection
\cite{zivkovic2004improved,rother2004grabcut,genovese2013fpga,
genovese2010opencv}. GMM parameters are typically estimated
using the Expectation–Maximisation (EM) algorithm, which
iteratively updates mixture means, variances, and component
weights. However, EM requires repeated evaluation of
exponential, logarithmic, and division operations, making
real-time online training computationally expensive.

FPGAs have been widely explored for accelerating GMM inference
due to their fine-grained parallelism and deep pipelining.
Prior work has demonstrated efficient hardware pipelines for
high-throughput likelihood evaluation
\cite{genovese2013fpga,he2017fully,kashiwagi2024fpga}.
However, these accelerators assume parameters are trained
offline and remain fixed during operation. This creates 
a fundamental limitation for long-lived edge systems, where 
scene statistics evolve continuously and retraining may be 
impossible due to bandwidth, power, or latency constraints. 
Consequently, they cannot adapt when the scene distribution changes.

To address this limitation, we propose
\textbf{MAGMA} ,
a fully synthesizable FPGA architecture that performs
concurrent GMM inference and online EM parameter
adaptation directly from a streaming pixel input.
MAGMA combines a pipelined inference datapath with a
background online update engine that accumulates
responsibility statistics and periodically executes
the EM M-step without interrupting the pixel stream.
The main contributions of this work are:
\begin{itemize}
\item \textbf{Streaming Online GMM Architecture:}
A fully synthesizable FPGA design that performs
GMM inference and EM parameter updates concurrently
from a streaming pixel input without stalling
inference.

\item \textbf{Hardware-Friendly EM Operators:}
Fixed-point implementations of the transcendental
operations required by EM, including a range-reduced
Chebyshev exponential, a CLZ-based logarithm
approximation, and a sequential divider.

\item \textbf{Robust Online Adaptation Mechanisms:}
Spatial sampling, conditional parameter updates, and
cluster-death safeguards that enable stable online
learning under changing scene statistics.

\end{itemize}
The remainder of this paper is organised as follows.
Section~\ref{sec:background} introduces the GMM and EM
algorithms together with the fixed-point approximation
techniques used in the hardware implementation.
Section~\ref{sec:related} reviews prior FPGA-based GMM
accelerators and positions MAGMA within existing work.
Section~\ref{sec:arch} describes the proposed architecture.
Section~\ref{sec:eval} presents the experimental evaluation,
including convergence and dynamic adaptation experiments.
Section~\ref{sec:discussion} discusses the results and
outlines future work.
Finally, Section~\ref{sec:conclusion} concludes the paper.
\section{Background}
\label{sec:background}

\subsection{Gaussian Mixture Models and the EM Algorithm}

A $K$-component Gaussian Mixture Model represents a probability density as a weighted sum of Gaussian distributions:
\begin{equation}
p(\mathbf{x}) = \sum_{k=1}^{K} \pi_k \, \mathcal{N}\!\left(\mathbf{x};\,\boldsymbol{\mu}_k,\,\Sigma_k\right)
\label{eq:gmm}
\end{equation}
where $\pi_k$, $\boldsymbol{\mu}_k$, and $\Sigma_k$ are the mixing weight, mean vector, and covariance matrix of component $k$, with $\sum_k \pi_k = 1$. For an RGB pixel $\mathbf{x}=(r,g,b)^\top$, inference assigns $\mathbf{x}$ to the component maximising the log-likelihood score:
\begin{equation}
\mathcal{S}_k(\mathbf{x}) = \log\pi_k - \tfrac{1}{2}\log|\Sigma_k| - \tfrac{D}{2}\log 2\pi - \tfrac{1}{2}(\mathbf{x}-\boldsymbol{\mu}_k)^\top \Sigma_k^{-1}(\mathbf{x}-\boldsymbol{\mu}_k)
\label{eq:score}
\end{equation}
so that $k^* = \arg\max_k\,\mathcal{S}_k(\mathbf{x})$. 
Under the isotropic covariance assumption $\Sigma_k = \sigma_k^2\mathbf{I}$, the quadratic term reduces to a scaled squared Euclidean distance:
\begin{equation}
d_k(\mathbf{x}) = \frac{\|\mathbf{x}-\boldsymbol{\mu}_k\|^2}{\sigma_k^2}
\label{eq:dist}
\end{equation}
eliminating matrix inversion and reducing inference to multiply-accumulate operations directly suitable for fixed-point hardware. Crucially, MAP classification via $k^* = \arg\max_k \mathcal{S}_k(\mathbf{x})$ requires \emph{no exponential evaluation} only subtraction and comparison. The exponential arises only in the online parameter update, described next.

For each pixel and each mixture component, online EM requires exponential evaluation, responsibility normalization, and parameter accumulation. These operations dominate runtime and motivate dedicated hardware acceleration.

GMM parameters are estimated via the Expectation-Maximisation~(EM) algorithm~\cite{dempster1977maximum}, which iterates two steps until convergence.

\textbf{E-step.} Compute the soft responsibility of component $k$ for observation $\mathbf{x}_n$:
\begin{equation}
r_{nk} = \frac{\pi_k \exp\!\bigl(-\tfrac{1}{2}d_k(\mathbf{x}_n)\bigr)}{\displaystyle\sum_{j=1}^{K} \pi_j \exp\!\bigl(-\tfrac{1}{2}d_j(\mathbf{x}_n)\bigr)}
\label{eq:estep}
\end{equation}
The exponential evaluations in~\eqref{eq:estep} must be computed for every pixel component pair at full pixel-stream throughput, making them the primary computational bottleneck in the hardware pipeline.

\textbf{M-step.} Update parameters by maximising the
expected complete-data log-likelihood. Define the
effective sample count $N_k = \sum_{n} r_{nk}$, then:
\begin{align}
\hat{\boldsymbol{\mu}}_k &= \frac{1}{N_k}\sum_{n} r_{nk}\,\mathbf{x}_n, \quad
\hat{\sigma}_k^2 = \frac{1}{D N_k}\sum_{n} r_{nk}\,\|\mathbf{x}_n - \hat{\boldsymbol{\mu}}_k\|^2
\label{eq:mu_sigma} \\[2pt]
\hat{\pi}_k &= N_k \Big/ {\textstyle\sum_j N_j}
\label{eq:pi}
\end{align}
These updates require division for normalisation and
logarithm evaluation for the derived constants
$\log\hat{\pi}_k$ and $\log|\hat{\Sigma}_k|$ used in
subsequent scoring~\eqref{eq:score}, both of which must
be realised in fixed-point hardware without
floating-point units.
In the online setting, updates are applied incrementally
after accumulating a mini-batch of $B$ pixels, enabling
the model to track non-stationary distributions without
storing the full dataset.

\subsection{Fixed-Point Arithmetic}

All internal signals in MAGMA use signed two's-complement fixed-point representation. A value in $Q_{m.n}$ format allocates $m$ bits to the integer part and $n$ bits to the fractional part, representing the value $v = b/2^n$ where $b$ is the stored integer. Multiplication of a $Q_{m_1.n_1}$ by a $Q_{m_2.n_2}$ value produces a $Q_{(m_1+m_2).(n_1+n_2)}$ result; restoring the target format requires a right-shift by $n_1{+}n_2$ bits, which costs no arithmetic logic in hardware. Similarly, normalising an 8-bit pixel value to $[0,1)$ is free: the integer is reinterpreted as a $Q_{0.8}$ fraction by prepending a zero integer bit, equivalent to dividing by 256 at zero hardware cost.
\begin{figure}[t]
    \centering
    \includegraphics[width=\columnwidth]{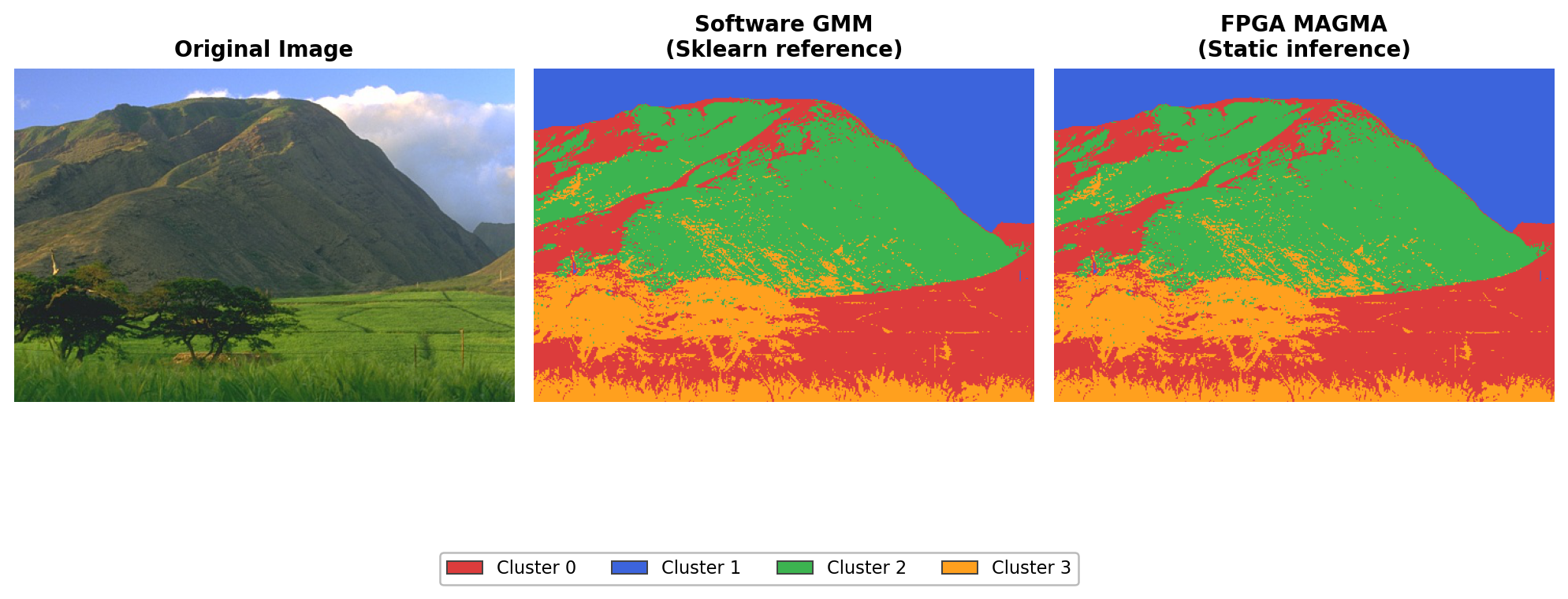}
    \caption{Cluster assignments after initialisation with parameters
learnt offline by a software GMM. Left: original RGB image.
Centre: software GMM reference (Scikit-learn). Right: MAGMA
FPGA static inference using the same quantised parameters.
The hardware pipeline achieves 99.5\,\% pixel accuracy
against the software reference, confirming that the
fixed-point quantisation preserves classification quality
at deployment. When trained directly using the online
update engine without offline initialisation, the resulting
clustering achieves a mean Intersection-over-Union (IoU)
score of 0.69 on the same image after 10 iterations.}
    \label{fig:comparison}
\end{figure}

\subsection{Function Approximation}

\subsubsection{Argument Reduction}

Polynomial approximations achieve acceptable accuracy only over a small interval. Argument reduction maps an arbitrary input to this interval using an exact algebraic identity, with the full result reconstructed afterwards at negligible cost.

For the exponential function, the identity:
\begin{equation}
\exp(x) = 2^n \cdot \exp(f), \quad n = \left\lfloor \frac{x}{\ln 2} \right\rfloor, \quad f = x - n\ln 2 \;\in\; [0,\ln 2)
\label{eq:range_red}
\end{equation}
restricts the polynomial argument to the compact interval $[0,\ln 2) \approx [0,\,0.693)$. The range of $f$ follows directly from the floor definition: $n \leq x/\ln 2 < n{+}1$ implies $0 \leq f < \ln 2$. The integer part $n$ is extracted by a fixed-point multiply and floor operation; reconstruction via $2^n$ is a single arithmetic shift, costing no DSP resources.

For the natural logarithm, the position of the most-significant set bit, extracted in one cycle by a Count-Leading-Zeros~(CLZ) operation, provides the integer part:
\begin{equation}
\ln(x) = \bigl(W - 1 - \mathrm{clz}(x)\bigr)\ln 2 + \ln\!\left(\frac{x}{2^{W-1-\mathrm{clz}(x)}}\right)
\label{eq:ln_rr}
\end{equation}
where $W$ is the word width and the second term is a fractional correction evaluated over the normalised mantissa $m \in [1,2)$ using a polynomial approximation.

\subsubsection{Chebyshev vs.\ Taylor Approximation}

Once the argument is reduced to a compact interval, the residual function is approximated by a polynomial. The classical Taylor series expanded around a fixed point $x_0$:
\begin{equation}
f(x) \approx \sum_{i=0}^{N} \frac{f^{(i)}(x_0)}{i!}(x-x_0)^i
\label{eq:taylor}
\end{equation}
minimises approximation error only at $x_0$; the error grows toward the interval boundaries, making Taylor series poorly suited to hardware where the full reduced interval must be covered uniformly.

Chebyshev~(Minimax) approximation instead minimises the \emph{maximum} absolute error over the entire target interval $[a,b]$:
\begin{equation}
p^*(x) = \arg\min_{p\,\in\,\mathcal{P}_N} \max_{x\,\in\,[a,b]}\bigl|f(x)-p(x)\bigr|
\label{eq:minimax}
\end{equation}
The optimal polynomial equioscillates between its worst-case error exactly $N{+}2$ times, guaranteeing the tightest possible worst-case bound for a given degree~\cite{cheney1966}. For $\exp(f)$ over $[0,\ln 2)$, a degree-3 Chebyshev polynomial achieves a maximum absolute error below $10^{-4}$, sufficient for 8-bit fixed-point classification. Coefficients are computed offline using the Remez algorithm~\cite{muller2006elementary} and embedded as fixed-point constants in the hardware.

\subsubsection{Horner Evaluation and DSP Mapping}

A degree-$N$ polynomial $p(x) = \sum_{i=0}^N c_i x^i$ requires $N$ multiplications and $N$ additions when evaluated naively. Horner's method rewrites the same polynomial as:
\begin{equation}
p(x) = c_0 + x\bigl(c_1 + x\bigl(c_2 + x(\cdots + x\,c_N)\bigr)\bigr)
\label{eq:horner}
\end{equation}
reducing evaluation to exactly $N$ multiply-accumulate~(MAC) operations with no intermediate powers of $x$. Each MAC maps directly to a single DSP48E1 block on Xilinx FPGAs, which natively executes $P \leftarrow A \cdot B + C$ in one clock cycle. A degree-3 Chebyshev polynomial therefore requires exactly three DSP blocks arranged as a four-stage pipeline:
\begin{align}
s_3 &= c_3 \label{eq:h1} \\
s_2 &= c_2 + x \cdot s_3 \label{eq:h2} \\
s_1 &= c_1 + x \cdot s_2 \label{eq:h3} \\
p   &= c_0 + x \cdot s_1 \label{eq:h4}
\end{align}
This Horner chain, preceded by the argument reduction of~\eqref{eq:range_red} and followed by the $2^n$ shift reconstruction, forms the basis of the exponential approximation unit described in Section~\ref{sec:arch}.
Building on these principles, we now describe the
hardware architecture of MAGMA that enables streaming
inference and concurrent online EM updates.
\section{Related Work}
\label{sec:related}

Several works have explored FPGA acceleration of Gaussian
Mixture Models and related probabilistic inference tasks,
yet none support concurrent online parameter adaptation
from a live pixel stream.
He~\textit{et al.}~\cite{he2017fully} present a fully-pipelined
EM-GMM design on Virtex-6 and Stratix-V FPGAs, achieving
over $200\times$ CPU speedup and $39\times$ over a Pascal
Titan-X GPU by restructuring the EM workflow and using a
fixed-point Gaussian PDF evaluation unit.
However, the M-step is explicitly offloaded to a host CPU
after each full dataset pass, providing no mechanism for
continuous adaptation to a streaming input.
Xu~\textit{et al.}~\cite{xu2021game} propose GAME, a
GMM-based mapping and navigation engine on a Xilinx ZCU102
achieving 59\,fps with $60\times$ GPU speedup using mixed
8/16-bit fixed-point quantisation; like He~\textit{et al.},
parameters are fixed before deployment.
Kashiwagi~\textit{et al.}~\cite{kashiwagi2024fpga}
implement an approximate GMM on FPGA for myoelectric
prosthetic control, achieving $40\times$ CPU speedup by
avoiding exponential evaluation entirely through log-space
classification; this eliminates the E-step responsibility
computation, making online EM updates impossible by design.
EhKan~\textit{et al.}~\cite{ehkan2011fpga} achieve
$90\times$ speedup for GMM-based speaker identification on
a Virtex-II FPGA using log-add approximations, targeting a
fixed offline-trained speaker model with no online update
capability.
Karthikeyan~\textit{et al.}~\cite{karthikeyan2025fpga}
accelerate image segmentation via ADMM on a Zynq
UltraScale+ MPSoC, achieving 9\,ms processing time; while
addressing hardware-accelerated segmentation, the design
uses a fundamentally different algorithm and does not
support probabilistic online model adaptation.

On the approximation side, prior FPGA GMM designs apply
degree-1 piecewise-linear functions without argument
reduction~\cite{genovese2013asic,he2017fully}, leaving
accuracy to degrade at interval boundaries.
MAGMA instead uses floor-based argument
reduction~\cite{muller2006elementary} combined with a
degree-3 Chebyshev polynomial~\cite{cheney1966} evaluated
via a Horner--DSP chain, achieving worst-case error below
$10^{-4}$ two orders of magnitude tighter.
More broadly, approximate computing exploits application-
level error resilience to trade precision for energy
savings~\cite{han2013approximate, mittal2016survey,
xu2015approximate}; MAGMA applies this to the GMM E-step,
where components with large Mahalanobis distance contribute
negligibly to responsibilities yet prior designs evaluate
every component at full precision indiscriminately.
Genovese and Napoli~\cite{genovese2013fpga,genovese2013asic}
implement the Stauffer--Grimson adaptive background model
on Virtex-6, achieving 47\,fps on 1080p video with
per-pixel online parameter updates entirely in hardware.
However, their update rule is a simplified running average
with a fixed learning rate $\alpha$ that requires no
exponential, division, or logarithm evaluation.
Assignment is based on a hard distance threshold rather
than soft probabilistic responsibilities, and each pixel
maintains an independent mixture model in block RAM.

To the best of our knowledge, MAGMA is the first
FPGA implementation of the \emph{complete} EM
algorithm , including soft responsibility computation
via fixed-point exponential, normalised M-step updates
via 48-bit division, and log-domain parameter
conversion , executing concurrently with inference
on a global GMM from a streaming pixel input, with
hardware guards against the cluster-death failure mode
that arises uniquely in full fixed-point EM.

\section{Methodology}
\label{sec:arch}

\subsection{System Overview}

\begin{figure*}[t]
\centering
\includegraphics[width=1\textwidth]{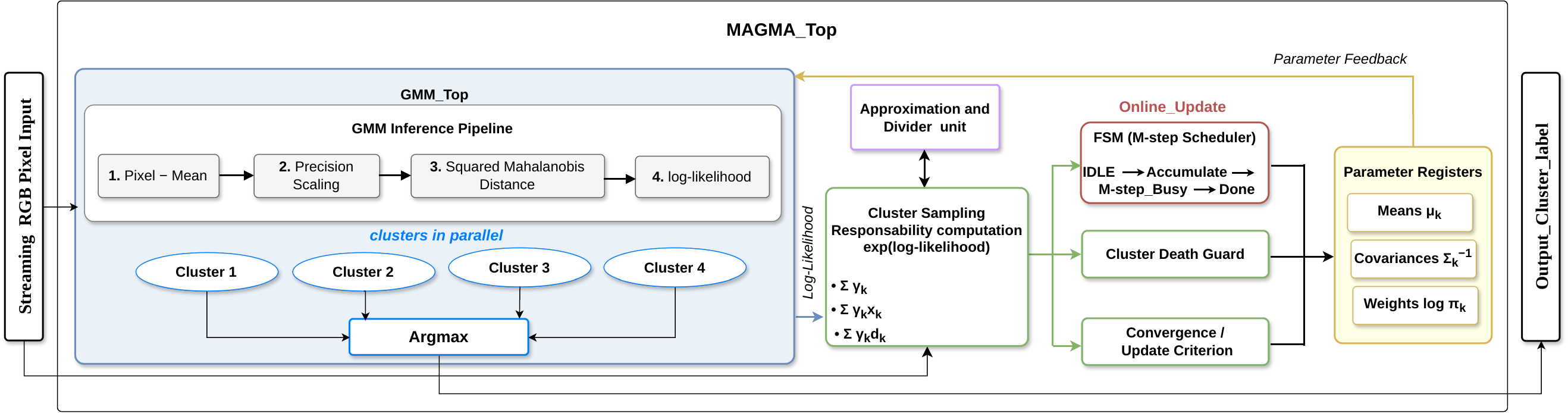}
\caption{Hardware architecture of the proposed MAGMA system. 
The inference pipeline evaluates per-cluster log-likelihoods 
for each pixel, while the Online Update engine performs 
periodic EM parameter updates without interrupting the 
streaming datapath.}
\label{fig:architecture}
\vspace{-4mm}
\end{figure*}

MAGMA is a fully synthesizable fixed-point FPGA architecture
that performs concurrent GMM inference and online EM parameter
adaptation directly from a streaming RGB pixel input.
As shown in Fig.~\ref{fig:architecture}, the system consists
of two primary modules instantiated within
\texttt{MAGMA\_Top}: the \texttt{GMM\_Top} inference pipeline
and the \texttt{Online\_Update} engine. These modules
communicate through an internal parameter bus that transfers
log-likelihood scores and updated model parameters.

For each incoming pixel, \texttt{GMM\_Top} evaluates the
log-likelihood under all Gaussian clusters and assigns the
pixel to the maximum a posteriori (MAP) cluster.
The resulting log-likelihoods are simultaneously forwarded
to \texttt{Online\_Update}, which accumulates responsibility
statistics over a configurable mini-batch and periodically
performs the EM M-step to update cluster means, variances,
and mixture weights. These updates occur entirely in
hardware and do not interrupt the inference datapath.

Cluster parameters are initially loaded through a
register write interface and subsequently updated
autonomously by the online learning engine.

\subsection{GMM Inference Pipeline}

The inference datapath instantiates $K$ parallel fixed-point
log-likelihood scoring units, one per cluster. Each unit evaluates

\begin{equation}
\mathcal{S}_k =
\log\hat{\pi}_k
- \tfrac{D}{2}\ln 2\pi
- \tfrac{1}{2}\texttt{log\_det}_k
- \frac{\|\mathbf{x}-\boldsymbol{\mu}_k\|^2}{\hat{\sigma}_k^2}
\label{eq:score}
\end{equation}

using a four-stage pipeline.

Stage~1 computes the signed pixel–mean differences
$\Delta_c = x_c - \mu_{k,c}$ for each colour channel.
Stage~2 scales these differences by the isotropic inverse
covariance \texttt{inv\_cov}$_k$, an 8-bit quantity representing
$\hat{\sigma}_k^{-2}$ scaled by the constant
\texttt{SCALE\_INV}.
Stage~3 accumulates the scaled Mahalanobis distance across all
three channels.
Stage~4 combines the resulting distance with the stored
log-determinant and log-prior to produce the log-likelihood
$\mathcal{S}_k$ in the fixed-point domain
.

A shared \texttt{Argmax} unit selects the highest-scoring cluster,
producing a MAP classification label with an end-to-end latency
of five cycles. All cluster parameters are stored in registers
and updated atomically between M-steps, ensuring that inference
never stalls during parameter adaptation.

\subsection{Exponential Approximation}

The E-step responsibility
$r_{nk} \propto \exp(\mathcal{S}_k)$ requires efficient
evaluation of the exponential function in hardware.
MAGMA implements this operation using the pipelined
exponential approximation module, which combines argument reduction
with a low-degree polynomial approximation.

The design uses floor-based argument reduction to decompose the
input into an integer exponent
$n = \lfloor x / (Q_{\text{LOG}}\ln 2) \rfloor$
and a reduced residual
$f = x - n\ln 2$.
The reduced term is then evaluated using a degree-3 Chebyshev
polynomial implemented via Horner’s rule~\cite{cheney1966}.

This approach yields a worst-case absolute approximation error
below $10^{-4}$ over the interval $[0,\ln 2)$ while maintaining
a  fully pipelined implementation compatible with the
streaming inference datapath.

A configurable clamp threshold suppresses extremely small
responsibilities prior to exponentiation, preventing numerical
noise from destabilising cluster assignments.

\subsection{Online EM Update Engine}

The \texttt{Online\_Update} module performs the EM M-step in
fixed-point hardware using a four-state finite-state machine:

\[
\textsc{Idle} \rightarrow
\textsc{Accumulate} \rightarrow
\textsc{Mstep\_Busy} \rightarrow
\textsc{Done}.
\]

During \textsc{Accumulate}, the engine collects soft statistics
from the streaming pixel data. When a sufficient number of
samples has been collected, the FSM transitions to
\textsc{Mstep\_Busy}, where updated model parameters are
computed. The new parameters are written atomically to the
register file during the \textsc{Done} state before the engine
returns to accumulation.

\subsubsection{Accumulation (E-step statistics)}

During \textsc{Accumulate}, the module does not process every
incoming pixel. High-resolution image streams may contain
hundreds of thousands of pixels per frame, and accumulating all
pixels unconditionally would cause the soft-count statistics to
grow extremely rapidly.

Let $P$ denote the number of pixels in a frame and assume
responsibilities are represented with 24 fractional bits.
Even moderate responsibilities lead to an accumulated count
$N_k \propto P \times 2^{24}$. For typical image sizes,
this value approaches $10^{12}$ after only a modest number of
updates, consuming a large portion of the 48-bit accumulator
range and risking overflow.

A second issue arises from spatial bias. If the M-step is
triggered mid-frame, the accumulated statistics reflect only
the region of the image processed most recently. This causes
cluster parameters to drift toward local colour distributions
rather than representing the global scene.

To address both issues, \texttt{Online\_Update} performs uniform
spatial subsampling using a modulo counter that accepts one
pixel every \texttt{sample\_modulus} cycles (default: every
40th pixel). This produces on the order of $10^4$ samples per frame for
typical video resolutions, while ensuring that each mini-batch
is drawn uniformly from across the image. The sampling rate is
controlled by the \texttt{sample\_modulus} parameter and can be
adjusted to accommodate different input resolutions without
modifying the hardware architecture.
\begin{figure*}[t]
\centering
\includegraphics[height=7cm]{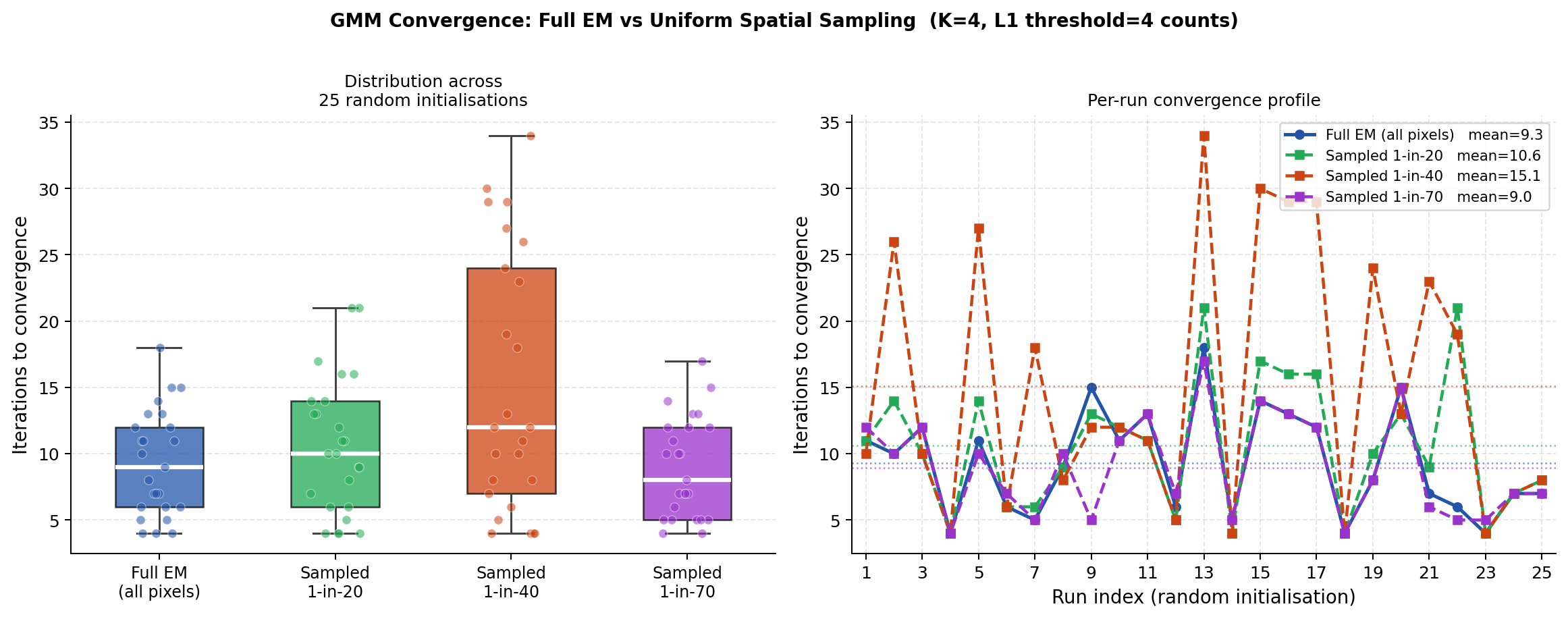}
\caption{Iterations required for EM convergence under different
spatial sampling rates. Even with subsampling (e.g., 1-in-40
pixels), the algorithm converges reliably, although more
iterations may be required compared to full-frame EM updates.
This confirms that uniform spatial sampling preserves
convergence while reducing the number of pixels processed
per update.}
\label{fig:convergence}
\vspace{-4mm}
\end{figure*}

The total number of accepted samples is bounded by
\texttt{batch\_size}, which decouples the M-step trigger from
frame boundaries and allows the update rate to be tuned
independently of image resolution.

For each accepted pixel, the following statistics are accumulated
for every cluster $k$:

\begin{align}
N_k     &\mathrel{+}= e_k, \\
S_{k,c} &\mathrel{+}= e_k x_c, \quad c \in \{R,G,B\}, \\
Q_k     &\mathrel{+}= e_k \|\mathbf{x}-\boldsymbol{\mu}_k\|^2,
\end{align}

where $e_k = \exp(\mathcal{S}_k)$ represents the unnormalised
responsibility in $Q24$ format.
The 48-bit accumulators support up to $B=10\,000$
soft-weighted samples without overflow.

Pixels for which all responsibilities evaluate to zero are
discarded to prevent stale zero-valued products from corrupting
the accumulated statistics.

\paragraph{Conditional Update Criterion}

To avoid unnecessary parameter updates after convergence,
MAGMA applies a conditional update rule. After each M-step,
the engine computes the L1 shift of the cluster means relative
to the previous parameters;that is  sum of absolute channel-wise
mean differences for each cluster:
\begin{equation}
\Delta_k = \sum_{c \in \{R,G,B\}}
|\hat{\mu}_{k,c} - \mu_{k,c}|
\label{eq:l1shift}
\end{equation}
If the maximum shift $\Delta_{\max} = \max_k \Delta_k$
falls below \texttt{converge\_thresh}, the newly computed
parameters are retained internally but not committed to
the inference parameter registers.
The accumulation statistics are reset and a new
accumulation phase begins, allowing the online engine
to continue monitoring the data stream for genuine
distribution changes.

\paragraph{Parameter Commit}

If the convergence condition is not satisfied, the newly
computed parameters are atomically written to the parameter
registers used by the inference pipeline.

\paragraph{ Cluster Death and Hardware Robustness Guards}
MAGMA implements three hardware guards that together
prevent numerical failure during online fixed-point EM.

\textbf{Guard~1 : Variance floor.}
After each M-step, $\hat{\sigma}_k^2$ is clamped to
a minimum of \texttt{MIN\_SIGMA2}$=64$ 
This prevents the variance collapse cycle in which a
dominant cluster shrinks its variance, attracts more
assignments, shrinks further, and eventually monopolises
the entire pixel distribution.

\textbf{Guard~2 : Cluster death check.}
Before issuing any division in the M-step, the engine
checks $N_k$.
If $N_k < 256$, the division is bypassed and the
variance floor is applied directly, preserving a valid
\texttt{inv\_cov} and keeping the cluster active.
Without this guard, $N_k \approx 0$ forces
\texttt{inv\_cov}$_k = 0$, collapsing the Mahalanobis
distance for all pixels and causing the affected cluster
to dominate all subsequent assignments --- a cascading
failure that reduced pixel accuracy from 96\% to 37\%
 in our experiments.

\textbf{Guard~3 : Conditional update criterion.}
After each M-step, the L1 shift of all cluster means
is computed.
If the maximum shift across all clusters falls below
\texttt{converge\_thresh}, the newly computed parameters
are retained internally but not committed to the
inference register file.
This prevents unnecessary parameter drift on stationary
scenes while allowing the engine to continue monitoring
the data stream for genuine distribution changes.
\begin{figure}[t]
\centering
\includegraphics[width=\columnwidth]{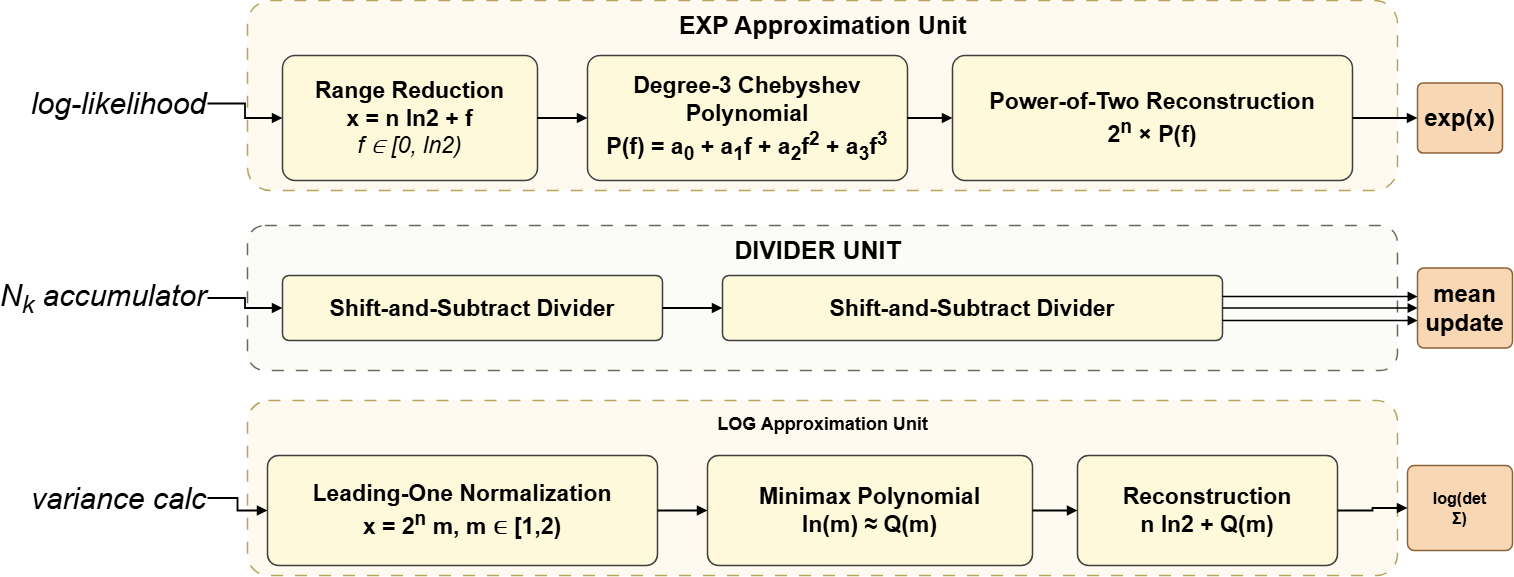}
\caption{Hardware datapath used for transcendental function approximation in MAGMA. 
The exponential unit uses range reduction followed by a third-order Chebyshev polynomial 
and power-of-two reconstruction. The logarithm unit uses leading-one normalization and 
a minimax polynomial approximation. Division operations required for the M-step are 
performed using a shift-and-subtract divider.}
\label{fig:approx}
\vspace{-4mm}
\end{figure}
\subsection{Fixed-Point Quantisation}
All log-domain quantities in the inference pipeline are
represented using $Q_{2.6}$ format
($Q_{\text{LOG}}{=}6$ fractional bits).
The logarithm approximation unit computes at $Q_{16}$
precision internally (Table~\ref{tab:approx}); the
result is truncated to $Q_{2.6}$ before commitment to
the parameter registers.

The fixed-point formats were selected
to provide sufficient dynamic range for EM parameter updates
while minimising datapath width and multiplier cost in the
FPGA implementation.

Table~\ref{tab:approx} summarises the transcendental function
approximations used in MAGMA.

\begin{table*}[t]
\centering
\caption{Transcendental function approximations used in MAGMA}
\label{tab:approx}
\begin{tabular}{llll}
\hline
\textbf{Function} & \textbf{Range Reduction} & \textbf{Approximation} & \textbf{Output Format} \\
\hline
$\exp(x)$ &
$x = n\ln2 + f$, $f \in [0,\ln2)$ &
Degree-3 Chebyshev polynomial &
Q8.24 \\

$\ln(x)$ &
$x = 2^n m$, $m \in [1,2)$ &
Quadratic minimax polynomial &
Q16 \\

Division &
None &
Shift-and-subtract divider &
48-bit \\

\hline
\end{tabular}
\vspace{-4mm}
\end{table*}

\section{Experimental Evaluation}
\label{sec:eval}

\subsection{Experimental Setup}

To evaluate the performance of the proposed MAGMA
architecture, experiments are conducted across two
complementary axes.
First, the convergence behaviour of the online EM
algorithm is characterised under varying spatial
sampling rates to validate that the hardware subsampling
strategy does not compromise learning stability Fig.~\ref{fig:convergence}.
Second, the online adaptation capability is evaluated
under a controlled non-stationary scene to assess
whether the architecture can track evolving scene
statistics and maintain numerical stability through
the hardware cluster-death guards.
All experiments use a $600 \times 398$ RGB test image
as the base scene with $K{=}4$ Gaussian clusters.
Three conditions are compared throughout:
\textbf{MAGMA  + guards} : The full synthesisable
hardware pipeline with variance floor and cluster death
guard active;
\textbf{Static baseline} : The same initial parameters
applied without any online update, modelling a
conventional fixed-parameter GMM classifier; and
\textbf{MAGMA  + no guards} : A MAGMA update logic with no guards 

\subsection{Convergence Under Spatial Subsampling}\label{sec:converge}
To validate the spatial sampling strategy, the EM training
procedure was executed on multiple natural images drawn from the
BSDS500 dataset~\cite{arbelaez2010contour}  across 25 independent random parameter initialisations
while varying the sampling rate: full-frame updates
and uniform subsampling at 1-in-20, 1-in-40, and
1-in-70 pixels.
Convergence was declared when the maximum L1 shift of
any cluster mean fell below a threshold of 4 counts
across consecutive M-step updates.
As shown in Fig.~\ref{fig:convergence}, the EM algorithm
converges reliably under all tested sampling rates.
Full-batch EM converged in a mean of 9.3 iterations
.
The hardware default of 1-in-40 required a mean of
15.1 iterations ($\sigma{=}9.6$), while 1-in-20 and
1-in-70 required 10.6 and 9.0 iterations respectively.
Critically, all 25 runs converged without exception
in every configuration.
The 1-in-40 rate incurs a median overhead of only
3 additional iterations while processing $40\times$
fewer pixels per update, providing a principled rather
than heuristic justification for the sampling
architecture: the $40\times$ reduction in per-update
sample volume does not significantly compromise the convergence
.

\subsection{Dynamic Scene Adaptation}
Section~\ref{sec:converge} already established MAGMA's
correctness on a \emph{static} scene: repeatedly presenting the
online engine with the same underlying image confirms that the
fixed-point EM update reliably converges to a stable clustering
across all tested spatial sampling rates (Fig.~\ref{fig:convergence}).
What this convergence result does not test is behaviour under a
\emph{genuinely non-stationary} scene --- one whose true underlying
colour distribution changes partway through, requiring the online
engine to track a moving target rather than simply settle onto a
fixed one. Evaluating this requires a sequence with a known,
controllable ground-truth distributional shift, which motivates the
synthetic construction described next.

To evaluate online adaptation under non-stationary con-
ditions, a synthetic 180-frame video sequence is generated
from the base image with four controlled phases: (1) gradual
illumination drift with a spectrally distinct cyan ellipse (15 percent 
of image area) introduced as a new object, (2) the object
fading out, (3) the object fully absent with a colour shift. The
cyan colour (0, 210, 180) is chosen to be spectrally alien to
the warm-toned base image, guaranteeing that it exclusively
occupies one of the four GMM clusters when present.
For each frame, a scikit-learn GMM re-fitted from scratch
on all pixels with full floating-point precision serves as the
per-frame reference. This shows the strongest possible baseline,
as the hardware is compared against the theoretically optimal
per-frame solution. Because cluster indices are arbitrary across
independent GMM fits, cluster identities are aligned using
Hungarian matching on the IoU matrix.
\begin{equation}
\text{IoU}(i,j) =
\frac{|C_i^{\text{HW}} \cap C_j^{\text{REF}}|}
     {|C_i^{\text{HW}} \cup C_j^{\text{REF}}|}
\label{eq:iou}
\end{equation}
Pixel accuracy after alignment is defined as:
\begin{equation}
\text{Accuracy} = \frac{1}{N}
\sum_{p=1}^{N}
\mathbf{1}\!\left(L^{\text{HW}}(p) =
L^{\text{REF}}(p)\right)
\label{eq:acc}
\end{equation}

We construct this sequence synthetically, rather than drawing
non-stationary scenes from an existing video benchmark such as
CDnet-2014~\cite{wang2014cdnet}, because CDnet was not built for the
task MAGMA performs. CDnet's ground truth encodes
\emph{semantic} foreground/background labels ,a human annotator's
judgement of which pixels belong to a real object of interest, such
as a person or vehicle , a category of structure that presupposes
object identity and motion. MAGMA, by contrast, performs
\emph{unsupervised global colour segmentation}: it partitions pixels
by colour-distributional similarity across the whole frame, with no
notion of object identity, motion, or semantic novelty. A global
colour model has no mechanism to represent CDnet's ground truth
directly, and scoring MAGMA against it would evaluate a task the
architecture was never designed to solve, rather than the property we
actually wish to test here: whether the online EM engine correctly
tracks a genuine, known change in the scene's colour distribution.
The synthetic construction above gives us exactly this , a
precisely-timed, colour-separable distributional shift with a known
ground truth, which is what a colour-clustering correctness test
requires and what no existing foreground/background benchmark
provides.

Table~\ref{tab:dynamic_results} summarises the results
by scene phase.

\begin{table*}[t]
\centering
\caption{Per-phase and overall segmentation performance
         on the 180-frame synthetic dynamic scene
         ($K{=}4$, $600\times398$ pixels).
         All metrics are Hungarian-matched against a
         per-frame scikit-learn reference.}
\label{tab:dynamic_results}
\setlength{\tabcolsep}{6pt}
\begin{tabular}{lcccccc}
\toprule
\multirow{2}{*}{\textbf{Phase}} &
\multicolumn{3}{c}{\textbf{Pixel Acc. (\%)}} &
\multicolumn{3}{c}{\textbf{Mean IoU}} \\
\cmidrule(lr){2-4}\cmidrule(lr){5-7}
 & \textbf{MAGMA+guards} & \textbf{MAGMA no guards} & \textbf{Static}
 & \textbf{MAGMA+guards} & \textbf{MAGMA no guards} & \textbf{Static} \\
\midrule
Object present (0--59)
  & 93.0          & 92.0          & \textbf{98.3}
  & 0.751         & 0.732         & \textbf{0.969} \\
Object fading  (60--119)
  & \textbf{77.0} & 76.5          & 71.9
  & \textbf{0.657}& 0.651         & 0.578 \\
Object absent  (120--179)
  & \textbf{74.5} & 68.8          & 69.0
  & \textbf{0.607}& 0.532         & 0.556 \\
\textbf{Overall mean}
  & \textbf{81.5} & 79.1          & 79.7
  & 0.672         & 0.638         & \textbf{0.701} \\
\bottomrule
\end{tabular}
\end{table*}

\subsection{FPGA Implementation}

MAGMA was implemented on an AMD Spartan-7 XC7S50-1CSGA324C FPGA
using Vivado 2020.2 with $K{=}4$ clusters and $D{=}3$ RGB
dimensions. Table~\ref{tab:resources} summarises the
post-implementation resource utilisation, while
Table~\ref{tab:hierarchy} and Table~\ref{tab:timing}
report the hierarchical breakdown and timing results,
respectively. The design meets timing at 74.49\,MHz and
consumes 7{,}779 LUTs, 4{,}237 FFs, and 91 DSPs without
requiring BRAM. At one pixel processed per cycle,
MAGMA achieves a throughput of 74.49\,Mpix/s.
Compared to the software baseline, the fully pipelined
inference datapath provides a $11.8\times$ speedup in
pixel classification throughput, while the hardware
EM update engine accelerates parameter adaptation
by up to $81\times$, enabling real-time inference
and online learning within a single FPGA pipeline.

\begin{table}[t]
\centering
\caption{Post-implementation resource utilisation of MAGMA
         on AMD Spartan-7 XC7S50 ($K{=}4$ clusters).}
\label{tab:resources}
\begin{tabular}{lrrr}
\hline
\textbf{Resource} & \textbf{Used} & \textbf{Available}
                  & \textbf{Util.\ (\%)} \\
\hline
LUT        & 7{,}779  & 32{,}600 & 23.86 \\
LUTRAM     &    63    &  9{,}600 &  0.66 \\
FF         & 4{,}237  & 65{,}200 &  6.50 \\
DSP        &    91    &    120   & 75.83 \\
\hline
\end{tabular}
\end{table}

\begin{table}[t]
\centering
\caption{Post-synthesis hierarchical resource breakdown.}
\label{tab:hierarchy}
\begin{tabular}{lrrrr}
\hline
\textbf{Module} & \textbf{LUT} & \textbf{LUTRAM}
               & \textbf{FF} & \textbf{DSP} \\
\hline
MAGMA Top           & 8{,}018 & 63 & 4{,}235 & 91 \\
\quad Online Update & 7{,}110 & 62 & 3{,}520 & 79 \\
\quad GMM Inference &   752   &  1 &   692   & 12 \\
\hline
\end{tabular}
\vspace{-4mm}
\end{table}
The online adaptation engine accounts for the majority of resource utilization, consuming 79 of 91 DSP blocks, highlighting that adaptation rather than inference dominates hardware cost.

\begin{table}[t]
\centering
\caption{Post-implementation timing summary.}
\label{tab:timing}
\begin{tabular}{ll}
\hline
\textbf{Metric} & \textbf{Value} \\
\hline
Clock period constraint  & 14.000\,ns \\
Target frequency         & 71.43\,MHz \\
WNS                      & $+$0.576\,ns \\
WHS                      & $+$0.034\,ns \\
WPWS                     & $+$6.02\,ns \\
Critical-path delay      & 13.424\,ns \\
Implied $f_\text{max}$   & 74.49\,MHz \\
\hline
\end{tabular}
\vspace{-4mm}
\end{table}
\begin{table*}[t]
\centering
\caption{Comparison of MAGMA with prior FPGA GMM
         accelerators. N/A indicates metric not reported.
         $\dagger$~M-step not in hardware.
         $\ddagger$~Stauffer--Grimson running average;
         no exponential, division, or logarithm required.
         $\S$~Speedup reported over GPU,
         not CPU.}
\label{tab:comparison}
\begin{tabular}{lcccccccccc}
\toprule
\textbf{Work} &
\textbf{Device} &
\textbf{Fmax} &
\textbf{LUT} &
\textbf{DSP} &
\textbf{BRAM} &
\textbf{Online?} &
\textbf{M-step} &
\textbf{M-step} &
\textbf{Inf.} &
\textbf{Target} \\
& & \textbf{(MHz)} & & & &
& \textbf{in HW?}
& \textbf{Speedup}
& \textbf{Speedup}
& \textbf{Task} \\
\midrule

Genovese~\cite{genovese2013fpga}
  & Virtex-6    & 97.19  & 922   & 0   & N/A
  & Yes & Partial$^\ddagger$ & N/A & N/A
  & BG subtraction \\

He~\textit{et al.}~\cite{he2017fully}
  & Stratix-V   & 200--250    & N/A   & N/A & N/A
  & No  & No$^\dagger$  & N/A & $>$200$\times$
  & Batch EM \\

Xu~\textit{et al.}~\cite{xu2021game}
  & ZCU102      & 250    & N/A   & N/A & N/A
  & No  & No$^\dagger$  & N/A & 60$\times^\S$
  & Robot mapping \\

Kashiwagi~\textit{et al.}~\cite{kashiwagi2024fpga}
  & N/A         & N/A    & N/A   & N/A & N/A
  & No  & No$^\dagger$  & N/A & 40$\times$
  & Prosthetic control \\

EhKan~\textit{et al.}~\cite{ehkan2011fpga}
  & Virtex-II   & 48    & 34,193   & 68 & 30
  & No  & No$^\dagger$  & N/A & 90$\times$
  & Speaker ID \\

\midrule

\textbf{MAGMA (ours)}
  & \textbf{Spartan-7}
  & \textbf{74.49}
  & \textbf{7,779}
  & \textbf{91}
  & \textbf{0}
  & \textbf{Yes}
  & \textbf{Full EM}
  & \textbf{81}$\times$
  & \textbf{11.8}$\times$
  & \textbf{RGB segmentation} \\

\bottomrule
\end{tabular}
\end{table*}
%
Table~\ref{tab:comparison} shows that prior accelerators
report higher inference speedups (40--207$\times$),
but this comparison is not direct for three reasons:
prior designs target mid-range to high-end FPGAs at
125--250\,MHz versus MAGMA's commodity Spartan-7 at
74.49\,MHz; they dedicate all resources to inference
because the M-step is offloaded to a host CPU, whereas
MAGMA allocates 86.8\% of DSPs and 88.7\% of LUTs to
the online update engine
(Table~\ref{tab:hierarchy}); and speedup ratios depend
on each work's CPU baseline.
MAGMA's 11.8$\times$ inference speedup should therefore
be read not as a limitation but as evidence that full
online EM adaptation can be added to an FPGA GMM
pipeline with only modest inference overhead , while
being the only design that performs both inference and
complete EM adaptation in a single autonomous pipeline
with no host CPU dependency, no block RAM, and at
274\,mW total power.

\section{Discussion}
\label{sec:discussion}
The dynamic scene experiment exposes the fundamental
plasticity--stability trade-off in online adaptive systems.
Both the static baseline and MAGMA are initialised from the
same offline-trained parameters, so any divergence in
subsequent behaviour is attributable solely to whether the
model is permitted to adapt, not to differences in starting
conditions.
During the stable opening phase, the static baseline achieves
98.3\% accuracy because its frozen parameters match the
initial scene perfectly; MAGMA incurs a mild over-adaptation
penalty (93.0\% with guards, 92.0\% without) as the
convergence guard cannot fully suppress stochastic M-step
shifts even on an otherwise stationary scene.

Despite this early disadvantage, MAGMA+guards' overall mean
accuracy across the evaluated sequence exceeds the static
baseline (81.5\% vs.\ 79.7\%), driven by the phases where the
scene evolves (frames 60--179): MAGMA outperforms the static
baseline by 5--5.5 percentage points in accuracy, as the
static model wastes a cluster on the absent object while
MAGMA migrates it toward the dominant background. The picture
is reversed for mean IoU, where the static baseline retains a
higher aggregate score (0.701 vs.\ 0.672) owing to its large
advantage during the opening phase alone (0.969 vs.\ 0.751);
across both non-stationary phases, MAGMA matches or exceeds
static IoU. Removing hardware guards degrades both metrics
consistently (MAGMA no-guards: 79.1\% accuracy, 0.638 IoU),
confirming that the variance floor and cluster-death check
contribute measurable robustness rather than merely preventing
catastrophic failure.

The relevant criterion for an always-on system is not
time-averaged accuracy under ideal initialisation but how
reliably the model adapts when the scene departs from its
training conditions. Starting both methods from identical
trained parameters isolates this effect directly: MAGMA's
online update engine recovers accuracy as the scene drifts,
while the static baseline's advantage is confined entirely to
the window in which the live scene still matches its frozen
initialisation. Note that this asymmetry would trivially
reverse if the original scene composition were restored: since
the static baseline's parameters are never altered, its
accuracy would immediately return to its initial level the
moment the scene once again matches its frozen initialisation,
whereas MAGMA's online update engine, having overwritten the
object-specific cluster during the intervening adaptation, has
no such guarantee (Section~\ref{sec:limitation}). This
highlights that the comparison favours whichever method's
assumptions better match the deployment scenario: static
parameters are preferable when the scene is expected to return
to a known configuration, while online adaptation is
preferable when genuine, lasting distribution shift is
expected.

\subsection{Limitations and Future Work}
\label{sec:limitation}

MAGMA maintains a single, shared $K$-component GMM over the entire
frame's colour space rather than an independent GMM per pixel
location as used by classical adaptive background
models~\cite{stauffer1999adaptive,zivkovic2004improved}; this is what
permits its constant, resolution-independent parameter footprint and
zero-BRAM implementation (Table~\ref{tab:resources}), since a
per-pixel-location GMM would instead require on the order of $10^6$
stored Gaussian parameter sets for a $600\times398$ frame at $K=4$.
The cost of this choice is spatial locality: per-pixel GMMs learn
each coordinate's ``normal'' appearance from that coordinate's own
history, tolerating legitimately variable regions (e.g.\ swaying
foliage) while still flagging novelty there, whereas MAGMA's global
clusters cannot distinguish two spatially distant regions with
similar colour statistics ,the same limitation that motivates our
choice not to benchmark directly against CDnet-2014, as discussed in
Section~\ref{sec:eval}. This absence of spatial locality may
nonetheless be an advantage precisely where the per-pixel assumption
breaks down: Sajid et al.~\cite{sajid2016appearance} show that a
single global appearance model with no spatial constraint outperforms
several pixel-wise methods on CDnet-2014's PTZ category (75.41\%
F-measure vs.\ a prior best of 62.07\%), attributing this directly to
removing the coupling between detector and spatial location. This
lends external support to the hypothesis that MAGMA's architecture
may hold a comparable advantage under camera motion, though this
remains an architectural prediction for MAGMA's specific fixed-point,
cluster-based formulation rather than a verified result. A further
limitation observed during evaluation is that MAGMA cannot recover a
previously learned mode once its dedicated cluster has been
overwritten during adaptation: an object that vacates the scene and
later reappears is treated as genuinely new, and the affected cluster
must be re-learned from scratch, unlike the static baseline, which
never updates and therefore recovers its original accuracy
immediately upon the scene's return. This catastrophic forgetting
behaviour motivates
several directions for future work: a hardware cluster memory or
exponential moving average over historical parameters to preserve
previously learned modes; an adaptive sampling rate responsive to
scene dynamics; a training-enable signal; per-region spatial GMMs
that would combine the resource-efficiency of a shared cluster bank
with a degree of spatial locality; dynamic $K$ via cluster splitting
and merging; and a real-world validation built around
colour-distributional or segmentation-quality ground truth, such as
BSDS500~\cite{arbelaez2010contour}, rather than the object-level
annotation that CDnet-style benchmarks provide.
\section{Conclusion}
\label{sec:conclusion}

This paper presented MAGMA, a fully synthesisable
fixed-point FPGA architecture that performs concurrent
GMM inference and online EM parameter adaptation from
a live pixel stream without interrupting inference
throughput.
MAGMA demonstrates that online probabilistic learning
can coexist with streaming FPGA inference in a single
autonomous architecture.
Unlike prior hardware GMM accelerators that rely on
offline-trained frozen parameters, MAGMA integrates
a pipelined inference datapath with a background online
update engine that executes the complete M-step ,
including fixed-point exponential, logarithm, and
48-bit division ,entirely in hardware.
Hardware guards against the cluster-death failure mode
unique to online fixed-point EM ensure numerical
stability across dynamic sequences.
Implemented on an AMD Spartan-7 XC7S50 consuming
274\,mW, MAGMA achieves $81\times$ M-step speedup and
$11.8\times$ inference throughput speedup,
 occupying 7,779
LUTs, 91 DSPs, and zero block RAM, demonstrating that
full online GMM learning is achievable in hardware with
modest resources on a commodity edge device.
\balance
\bibliographystyle{IEEEtran}
\bibliography{ref}
\end{document}